\documentclass[aps,pre,reprint,amsmath,amssymb,floatfix,citeautoscript,superscriptaddress]{revtex4-1}
\usepackage{cancel}
\usepackage{amsmath}
\usepackage{physics}
\usepackage{graphicx}
\usepackage{amssymb}
\usepackage{amsthm}
\usepackage{bm}
\usepackage{threeparttable} 
\usepackage{dcolumn}
\newcolumntype{d}[1]{D{.}{.}{#1}}
\usepackage{braket}
\usepackage{booktabs}
\usepackage{ragged2e}
\usepackage{txfonts}
\usepackage{color}
\usepackage[usenames,dvipsnames]{xcolor}

\definecolor{myblue}{rgb}{0,0,1}
\usepackage[breaklinks=true,colorlinks=true,linkcolor=myblue,urlcolor=myblue,citecolor=myblue]{hyperref}

\newcommand{\vq}{{\bm{q}}}
\newcommand{\vk}{{\bm{k}}}
\newcommand{\mesh}[1]{\ensuremath{{#1}\times{#1}\times{#1}}}

\begin{document}

\title{Absorption Spectra of Solids from Periodic Equation-of-Motion Coupled-Cluster Theory}

\author{Xiao Wang}
\affiliation{Center for Computational Quantum Physics, Flatiron Institute, New York, New York 10010 USA}

\author{Timothy C. Berkelbach}
\email{tim.berkelbach@gmail.com}
\affiliation{Center for Computational Quantum Physics, Flatiron Institute, New York, New York 10010 USA}
\affiliation{Department of Chemistry, Columbia University, New York, New York 10027 USA}

\begin{abstract}
We present ab initio absorption spectra of six three-dimensional semiconductors and insulators calculated
using Gaussian-based periodic equation-of-motion coupled-cluster theory with single and double excitations
(EOM-CCSD).
The spectra are calculated efficiently by solving a system of linear equations at each frequency, giving
access to an energy range of tens of eV without explicit enumeration of excited states. We assess the impact
of Brillouin zone sampling, for which it is hard to achieve convergence due to the cost of EOM-CCSD. Although
our most converged spectra exhibit lineshapes that are in good agreement with experiment, they are uniformly
shifted to higher energies by about 1~eV. We tentatively attribute this discrepancy to a combination
of vibrational effects and the remaining electron correlation, i.e., triple excitations and above.
\end{abstract}

\maketitle

\section{Introduction}

Absorption spectroscopy is an important tool for studying the electronic
properties of materials.
For semiconductors and insulators, the low energy part of the absorption spectrum is typically
dominated by excitonic effects, which originate from electron-hole interactions that may be 
screened by the other electrons.  
The most common computational methods currently used for simulating absorption spectra
are time-dependent density functional theory (TDDFT)~\cite{runge1984,reining2002,ullrich2016} and the Bethe-Salpeter equation
based on the GW approximation to the self-energy (GW-BSE)~\cite{sham1966,hanke1980,albrecht1998,rohlfing2000,onida2002}. 
In TDDFT, it has long been
recognized that the inclusion of nonlocal exchange is critical for the description of
excitons~\cite{Bruneval2006,Botti2007,Paier2008,izmaylov2008}, and promising recent work has applied
screened or dielectric-dependent
range-separated hybrids~\cite{Wing2019,Tal2020}.
The GW-BSE approach is based on time-dependent many-body perturbation theory and
typically includes screening at the level of the random-phase approximation. The
predictions are reasonably accurate when compared to experiments, although
implementation details, starting point dependence, and
the absence of finite-temperature or vibrational effects make a rigorous
evaluation challenging. For example, benchmark studies on 
molecules~\cite{bruneval2015,jacquemin2016,jacquemin2017,gui2018} have found
that the GW-BSE results depend strongly on the reference
functional and the optimal functionals are very different than those typically used
for solid-state calculations. 

Recent years have seen rapid development of wavefunction-based quantum chemistry techniques
for periodic solids~\cite{hirata2001,hirata2004,katagiri2005,gruneis2011,mcclain2017,gruber2018,zhang2019}.  
In the present context of neutral excitation energies, equation-of-motion 
coupled-cluster theory with single and double excitations (EOM-CCSD)
is a promising alternative to TDDFT or GW-BSE. For example, our group has
applied EOM-CCSD to
study plasmons in models of metals~\cite{Lewis2019,Lau2020}, as well as singly- and
doubly-excited states in a molecular crystal~\cite{Lewis2020}. Recently, the two
of us presented a systematic study of EOM-CCSD for a range of semiconductors and 
insulators, finding an accuracy of about 0.3~eV for the first singlet excitation
energy~\cite{Wang2020}. Although these preliminary results are encouraging, 
the optical response of solids is encoded
in the full energy-dependent absorption spectrum, which depends on all excited
states in the energy range of interest and their oscillator strengths. Here, we
extend our previous work and study the absorption spectra of semiconductors
and insulators predicted by EOM-CCSD. 

The remainder of the paper is organized as follows.  In Sec.~\ref{sec:theory},
we briefly describe the theory underlying the calculation of solid-state
absorption spectra with periodic EOM-CCSD. In Sec.~\ref{sec:comput}, we provide
computational details about the basis sets used, integral
evaluation, and $k$-point sampling.  In Sec.~\ref{sec:results}, we first
present and discuss our final EOM-CCSD optical absorption spectra for six
solids, before demonstrating detailed studies of the impact of various
approximations.  Finally, in Sec,~\ref{sec:conc}, we summarize our results and
conclude with future directions.

\section{Theory}
\label{sec:theory}

Within EOM-CC theory~\cite{emrich1981,emrich1981a,koch1990,stanton1993,bartlett2007,krylov2008,bartlett2012}, 
excited states with momentum $\vq$ are
given by
\begin{equation}
|\Psi(\vq)\rangle = \hat{R}(\vq) e^{\hat{T}} |\Phi_0\rangle
\end{equation}
where $\hat{T}$ creates momentum-conserving particle-hole excitations
and $\hat{R}$ creates particle-hole excitations with momentum~$\vq$.
The $\hat{T}$ operator is determined by the solution of the nonlinear
CC amplitude equations and the $\hat{R}$ operator is determined by a 
non-Hermitian matrix eigenvalue problem.
In crystals, the density of excited states prohibits their 
direct enumeration.
Instead, the absorption (or scattering) spectrum $S_\vq(\omega)$ can be obtained
directly at arbitrary frequency by using the solution to a system of linear
equations,
\begin{subequations}
\label{eq:spectrum_final}
\begin{align}
&S_{\vq}(\omega) = -\pi^{-1} \mathrm{Im} \langle \Phi_0 | (1 + \hat{\Lambda}) 
    \hat{\bar{\mu}}_\vq^\dagger |x_\vq(\omega) \rangle \\
\label{eq:spectrum_lineq}
&[\omega - (\hat{\bar{H}} - E_0) + i\eta] |x_\vq(\omega) \rangle 
    = \hat{\bar{\mu}}_\vq |\Phi_0 \rangle. 
\end{align}
\end{subequations}
where $\hat{\bar{O}} = e^{-\hat{T}} \hat{O} e^{\hat{T}}$ are similarity-transformed operators,
$\hat{\Lambda}$ is the deexcitation operator needed for expectation values
in CC theory, 
$\hat{\mu}_\vq = \sum_{cv\vk} \mu_{c\vk+\vq,v\vk} \hat{a}^\dagger_{c\vk+\vq} \hat{a}_{v\vk}$
is the transition operator with momentum $\vq$,
and $\eta$ is a numerical Lorentzian linewidth.

Here, we study the performance of EOM-CC with single and double
excitations (EOM-CCSD), i.e.~$\hat{T}=\hat{T}_1+\hat{T}_2$, 
$\hat{\Lambda} = \hat{\Lambda}_1 + \hat{\Lambda}_2$, and  $\hat{R}=\hat{R}_1+\hat{R}_2$, and focus on
absorption spectra with $\vq=0$.
For each frequency $\omega$, the cost of iteratively solving the system of 
linear equations~(Eq.~\ref{eq:spectrum_lineq}) scales as $O(N_k^4 N_o^2 N_v^4)$,
where $N_k$ is the number of $k$-points sampled
in the Brillouin zone and $N_o,N_v$ are the number of occupied
and virtual orbitals in the unit cell. 
In practice, the iterative solution converges slowly for some
values of $\omega$. Therefore, in this work we also test
and apply so-called partitioned EOM-CCSD~\cite{nooijen1995,stanton1995,gwaltney1996}, where the double
excitation block of the similarity transformed Hamiltonian is
approximated by a diagonal matrix of orbital energy differences.
This reduces the iterative scaling of the EOM step to 
$O(N_k^3 N_o^2 N_v^3)$.

As an alternative to EOM-CC, one can use the linear-response coupled cluster (LR-CC) theory 
to calculate 
excited-state properties. When no truncation is carried out in the excitation 
levels, LR-CC and EOM-CC both give exact results. At a truncated excitation level,
the methods yield identical excitation energies but different 
excited-state properties, such as transition dipole moments, and only properties
predicted by LR-CC are properly size extensive~\cite{kobayashi1994,koch1994}.
Although this finding calls into question the applicability of EOM-CCSD for solid-state
absorption spectra, the violation of size extensivity is
strongly mitigated when large basis sets are used~\cite{Caricato2009}.
In this work, we observe no problems associated with this deficiency of
EOM-CCSD for spectra, perhaps because of the near completeness of the basis set in
periodic solids.

\section{Computational Details}
\label{sec:comput}

The relatively high cost of periodic CCSD calculations makes it challenging to achieve
convergence to the complete basis set and thermodynamic limits. We have tested convergence with
respect to Brillouin zone sampling, basis sets, frozen orbitals, and the partioned EOM approximation,
which will be discussed in Sec.~\ref{ssec:approx}. 
Based on our studies, our final calculations presented here are performed in the
following way. We use GTH pseudopotentials~\cite{goedecker1996,hartwigsen1998} and the corresponding
polarized double-zeta basis set (DZVP)~\cite{vandevondele2005}.  Two-electron repulsion integrals
were treated by Gaussian density fitting with an even-tempered auxiliary basis 
(see ref~\citenum{sun2017} for more details).  In the CCSD calculations,
we correlate the highest four occupied and the lowest four virtual orbitals at
each $k$-point, while all of the other orbitals are frozen.  The partitioning
approximation is made to the similarity transformed Hamiltonian whereby the
dense doubles block is replaced by a diagonal matrix of orbital
energy differences~\cite{nooijen1995,stanton1995,gwaltney1996}. The Brillouin
zone was sampled with a uniform mesh of up to $N_k=5\times5\times5$ $k$-points.  
The $k$-point mesh is shifted to include
either the $\Gamma$ point or the random symmetry-breaking point 
$\vk=(0.11, 0.21, 0.31)$ (in fractions of the reciprocal lattice vectors).
Such random shifts have been previously shown to yield absorption spectra that
converge more quickly to the thermodynamic limit~\cite{ahmadpourmonazam2013}.
Lastly, we separately treat the convergence
of the first excitation energy and extrapolate to the thermodynamic limit by
assuming finite-size errors that scale as $N_k^{-1}$, as discussed in our previous 
work~\cite{Wang2020}. We then rigidly shift the entire absorption spectrum by
this finite-size correction, which is 0.1--0.4~eV for the materials and $k$-point
meshes considered here.
All calculations were performed with PySCF~\cite{sun2018,sun2020a}.

\section{Results and discussion}
\label{sec:results}

\subsection{EOM-CCSD absorption spectrum for six solids}

\begin{figure*}
	\centering
	\includegraphics[width=\textwidth]{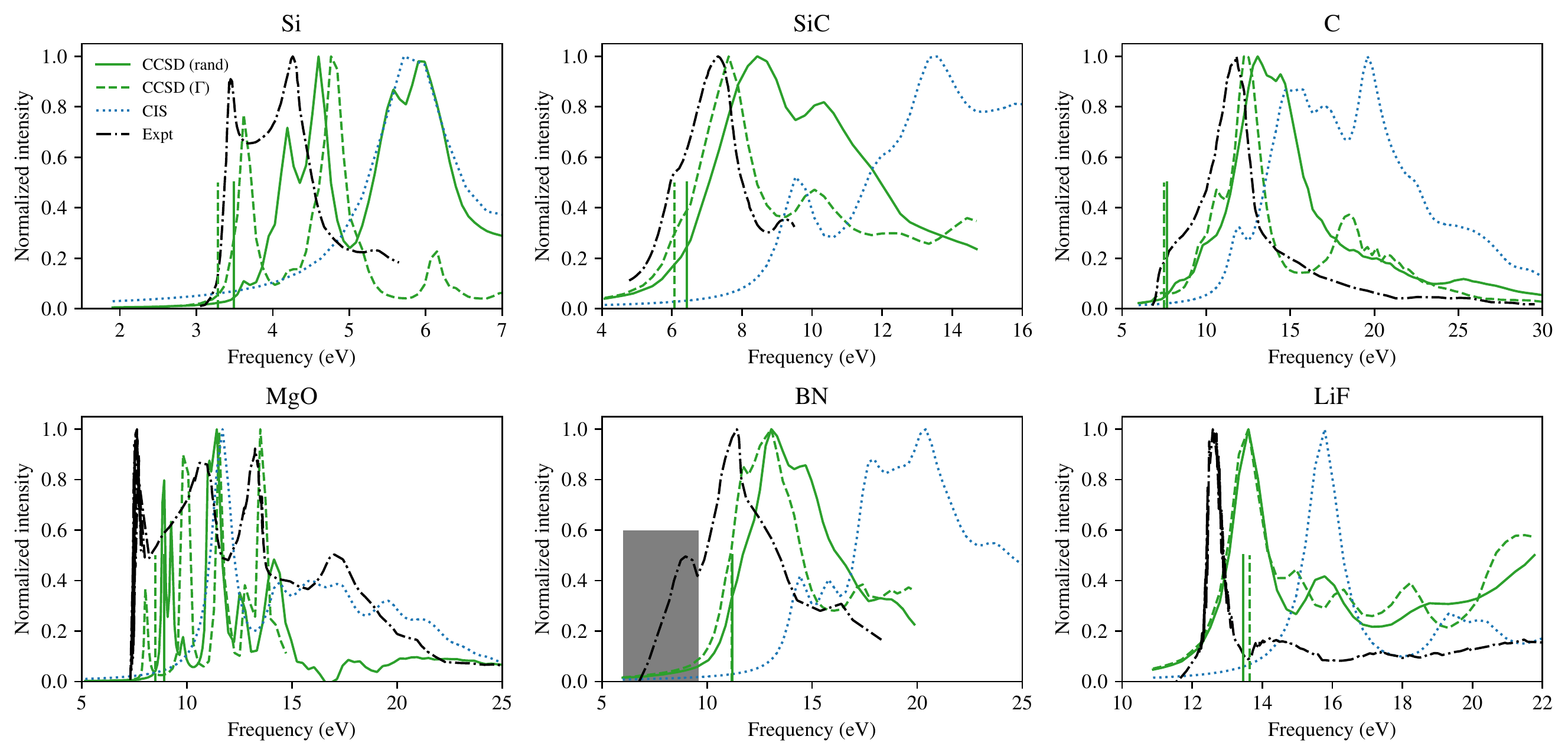}  
	\caption{Absorption spectra of Si, SiC, C, MgO, BN, and LiF in the DZVP basis set.
		A \mesh{5} $k$-point mesh is used for all CCSD spectra (green) and a \mesh{7} $k$-point
		mesh for all CIS spectra (blue).
		CCSD spectra are shown with $k$-point meshes that are shifted to include the
		$\Gamma$ point (dashed green) and a random, symmetry-breaking $k$-point
		(solid green). The corresponding EOM-CCSD first excitation energies are 
		indicated by green vertical lines.
		A broadening of $\eta=0.54$~eV is used in all calculations except for silicon
		and MgO where a smaller broadening of $\eta=0.08$~eV is used to resolve the
		sharp peaks. For BN, the shaded region of the experimental spectrum should be 
		ignored as it has been attributed to defects and polymorphism~\cite{tararan2018}.
	}
	\label{fig:spectra}
\end{figure*}

In Fig.~\ref{fig:spectra}, we show our best and final results for the EOM-CCSD
absorption spectra of six three-dimensional semiconducting and insulating materials:
Si, SiC, C, MgO, BN, and LiF. The experimental lattice constants were used for all systems: 
Si (5.431~\AA), SiC (4.350~\AA), C (3.567~\AA), MgO (4.213~\AA), BN (3.615~\AA), and LiF (4.035~\AA). 
Spectra were obtained using a $5\times 5\times 5$ $k$-point mesh;
in order to give some sense of possible finite-size
errors, we show EOM-CCSD results obtained with the two $k$-point shifts mentioned above. 
Our EOM-CCSD absorption spectra are compared to experimental ones and
to those obtained by configuration interaction with single excitations (CIS),
which was performed with a denser $7\times 7\times 7$ $k$-point mesh.

As seen in Fig.~\ref{fig:spectra}, the EOM-CCSD spectra are in reasonably good agreement with experiment.
Different $k$-point shifts give similar spectra for large gap insulators (like LiF and BN) and 
different spectra for smaller gap semiconductors (like Si and SiC), whose main features are shifted
from one another by as much as 1~eV. When compared to experiment, the EOM-CCSD spectra are shifted
to higher energies by about 1~eV, but otherwise have very similar lineshapes, indicating an accurate
description of excitonic interactions and concomitant redistribution of spectral weight. By comparison,
CIS spectra massively overestimate the excitation energies of solids by 3~eV or more, as shown in
our previous work~\cite{Wang2020}, and often have qualitatively incorrect spectral structure.
Because Hartree-Fock-based CIS is identical to unscreened GW-BSE, these results emphasize
the well-known importance of screening, especially in semiconductors.

We believe that the shift to higher energies that is exhibited by EOM-CCSD when compared 
to experiment is mostly attributable to the missing correlation due to the neglect of 
triple (and higher) excitations and the absence of vibrational and finite-temperature effects,
which are of course present in experiments and absent in the calculations. With regards to
electron correlation, it is interesting to note that our previous work, which
did not study spectra, found that EOM-CCSD overestimated the first excitation
energy by about 0.3~eV, which is noticably smaller than the 
deviations seen in the spectra in Fig.~\ref{fig:spectra}.  This discrepancy (i.e.~overestimation
by 0.3~eV versus 1~eV or more) is because
the first excited state, especially in indirect gap materials, is typically
weakly absorbing and contributes to the gradual onset of absorption. However,
experimentally reported first excitation energies are typically those of
spectral peaks or intense features, which occur at higher energies than the
onset of absorption. In Fig.~\ref{fig:spectra}, the green vertical lines indicate the calculated
first excitation energies, which are corrected for finite-size effects and
other approximations mentioned in Sec.~\ref{sec:comput}. 

In contrast to the apparent
differences in spectra, the use of differently shifted $k$-point meshes produce first excitation 
energies that agree reasonably well with each other, with a difference of 0.04--0.4~eV.
The first excitation energies differ from 
our previously reported values~\cite{Wang2020} (by 0.2~eV or less) due to a slightly different treatment of the 
finite-size effects in the current work, i.e.~(a) we extrapolate the data using a function of the form 
$E_\infty + aN_k^{-1}$,
(b) here, the $5\times 5\times 5$ result is included in extrapolation, and (c) the partitioning 
approximation is corrected by a constant shift deduced from the $3\times 3\times 3$ result.
Naturally, we believe that the comparison of spectra, as done here,
enables the most fair evaluation of the accuracy.  However, this overestimation of excited states
by 1~eV or more is significantly higher than the known performance of EOM-CCSD in molecules.
This difference is surprising, especially given that the excitonic states
contributing to absorption are all predominantly single-excitation in character and
that the EOM-CCSD polarizability has most of the diagrammatic content of the GW-BSE polarizability,
plus more~\cite{Lange2018,Berkelbach2018,Lewis2019,Lewis2020}.

Among the six solids in Fig.~\ref{fig:spectra}, a few show noticeable
differences between the EOM-CCSD and experimental spectra.
The worst agreement is for silicon, which has the smallest gap of all materials considered.
In its experimental spectrum, the main features are the two peaks at 3.5~eV and 4.3~eV with similar intensity.
While the CCSD spectra with both $k$-point shifts predict the position of the first peak reasonably well,
the randomly shifted $k$-point mesh severely underestimates its intensity 
relative to that of the higher-lying peaks.
In contrast, the $\Gamma$-inclusive $k$-point mesh correctly gives similar intensity 
for the two-peak structure, although the intensity between the two peaks is strongly underestimated.
We believe that the poor agreement between theory and experiment is due to the large remaining finite-size effects,
which are expected to be largest for this small-gap semiconductor.

\subsection{Approximations and error corrections}
\label{ssec:approx}

Finite-size errors of excited-state properties like absorption spectra
have been widely discussed in the TDDFT and GW-BSE literature~\cite{rohlfing2000,laskowski2005,fuchs2008,sander2015,Wing2019},
in part due to the relative maturity and low computational cost of these methods.
In constract, the finite-size errors of wavefunction-based methods such
as CCSD have been studied signficantly less, especially for spectra.
In the following, we will use diamond as an example to study the finite-size errors
of the spectra predicted by the EOM-CCSD.

\begin{figure}[t]
	\centering
	\includegraphics[width=\linewidth]{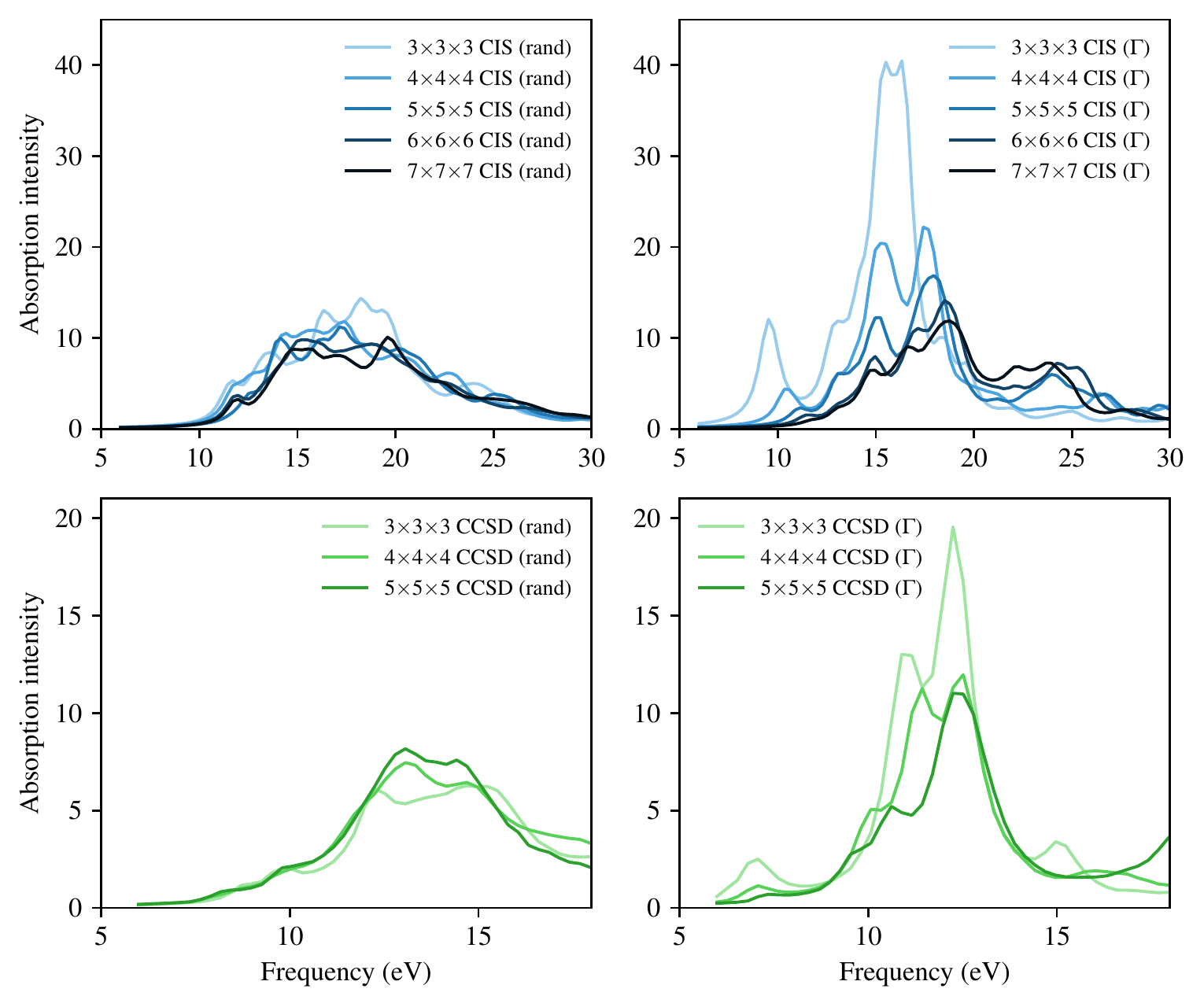}  
        \caption{Convergence of the CIS (top) and EOM-CCSD (bottom) spectra of diamond
            using various $k$-point shifts and sampling densities.}
	\label{fig:kcenter}
\end{figure}

As a warm-up to EOM-CCSD, we first consider CIS, which forms a minimal theory for 
electronic excited states in the condensed phase and is qualitatively comparable to TDDFT and GW-BSE. 
Importantly, the relatively low cost of CIS allows us to study the convergence with respect to 
Brillouin zone sampling up to relatively large $k$-point meshes.
In the upper panels of Fig.~\ref{fig:kcenter}, we show the CIS absorption spectra computed with various 
$k$-point meshes centered at $\Gamma$ (right column) or randomly shifted (left column), including up to \mesh{7} $k$-point meshes.
At low mesh densities (like \mesh{3}), the spectra computed with different $k$-point shifts 
show a large discrepancy in both peak positions and intensities.
This discrepancy is due to insufficient Brillouin zone sampling and largely depends on details of the
band structure.
As the mesh density is increased, the spectra converge to a similar result, but
the convergence is much more rapid with the randomly shifted $k$-point mesh.
Even for this insulator, the CIS spectra are not graphically converged with a
\mesh{7} mesh. This must be kept in mind when analyzing the EOM-CCSD spectra in
Fig.~\ref{fig:spectra}, which are limited to \mesh{5} meshes.

In the lower panels of Fig.~\ref{fig:kcenter}, the EOM-CCSD spectra of diamond with the same two $k$-point shifts are shown, for mesh
densities ranging from \mesh{3} to \mesh{5}.
As for CIS, we again see that the randomly shifted mesh provides significantly faster convergence towards the thermodynamic 
limit. In fact, the \mesh{4} and \mesh{5} are very similar and suggest semiquantitative convergence, especially at low
energies.

\begin{figure}[t]
	\centering
	\includegraphics[width=\columnwidth]{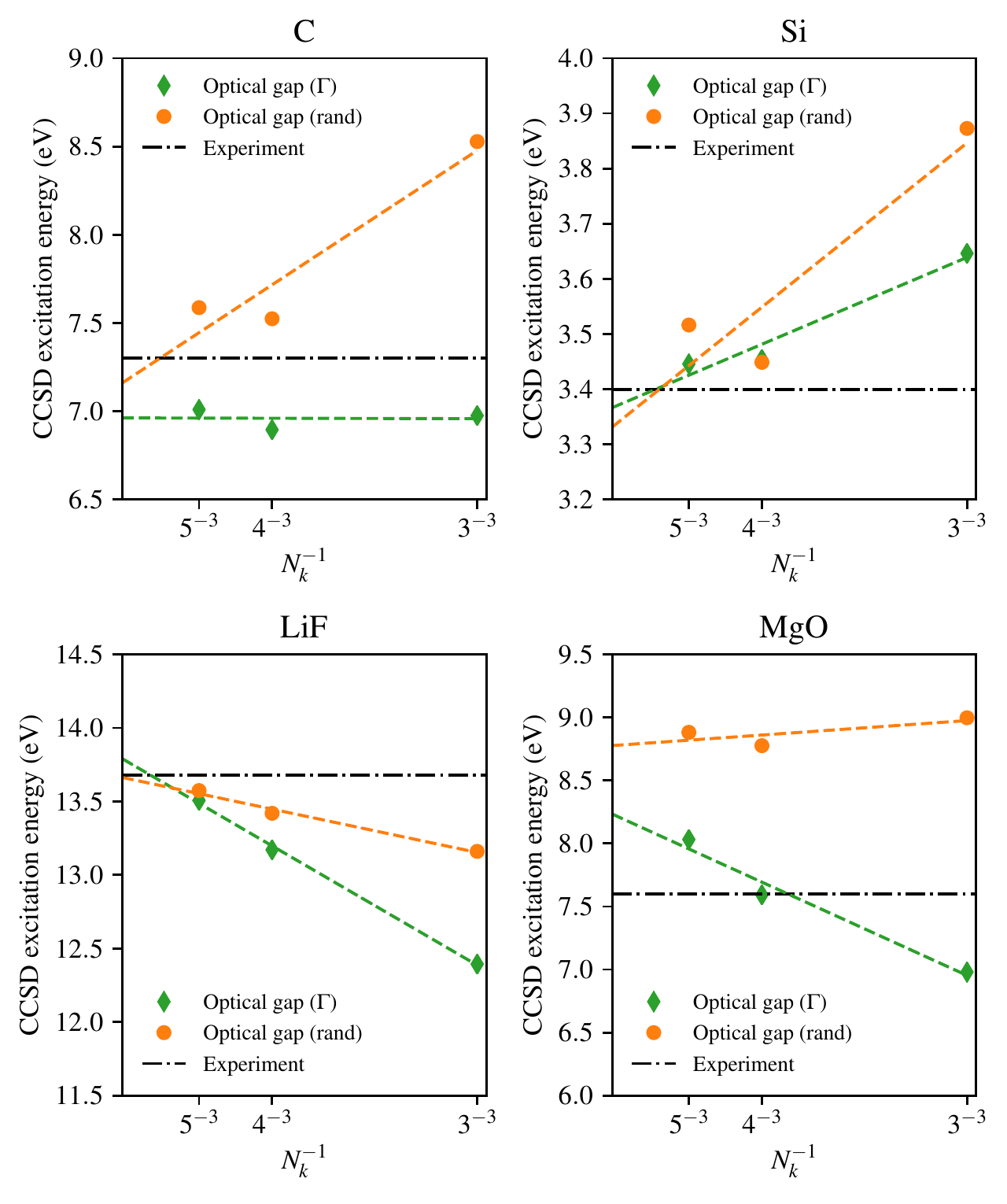}  
	\caption{Extrapolation of first excitation energies to the thermodynamic limit for C, Si, LiF, and MgO. Frozen virtual orbitals and partitioning are used in all cases.}
	\label{fig:extrap}
\end{figure}

In addition to the spectral intensities, the excitation energies also exhibit
large finite-size errors. These latter finite-size errors are simpler to remove
by extrapolation. Our final EOM-CCSD spectra shown in Fig.~\ref{fig:spectra}
have been rigidly shifted by the finite-size error of the first excitation
energy. This finite-size error is determined by extrapolation, assuming that the
finite-size error decays as $O(N_k^{-1})$.  Raw data and extrapolation fits are
shown in Fig.~\ref{fig:extrap} for four of the solids studied here. As expected,
we see that the convergence is erratic for indirect gap materials (C and Si) but
significantly smoother for direct gap materials (LiF and MgO). At the largest $k$-point
meshes, the finite-size errors are in the range of 0.1--0.4~eV.

\begin{figure}[t]
	\centering
		\includegraphics[width=\columnwidth]{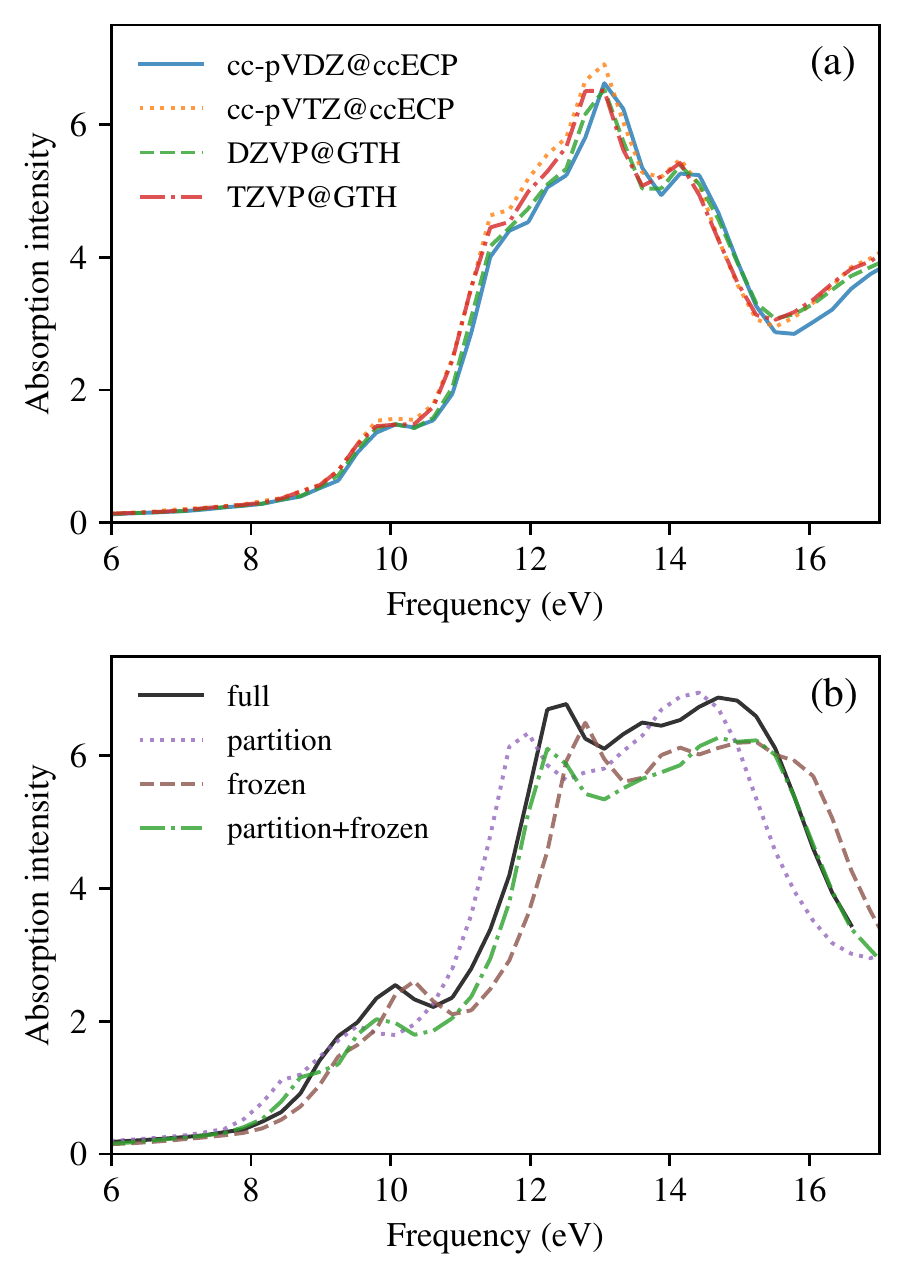}
        \caption{EOM-CCSD absorption spectra of diamond with various basis sets 
        	and approximations as indicated. A random $k$-point shift and 
        	a Lorentzian broadening of $\eta=0.54$~eV is used in all cases.
                (a) Comparison of EOM-CCSD spectra (without partitioning and
                with all orbitals correlated) using four different basis 
        	set/pseudopotential combinations as indicated. The Brillouin zone was sampled 
        	with a $2\times2\times2$ $k$-point mesh.
        	(b) Comparison of the full EOM-CCSD spectra and various approximations as
                indicated using the DZVP basis set and GTH pseudopotential.
                The Brillouin zone was sampled with a $3\times3\times3$
                $k$-point mesh.}
		\label{fig:approx} 
\end{figure}

Beyond the finite-size errors, we have studied the effects of three other approximations:
incomplete basis set, frozen orbitals, and the partitioning of EOM-CCSD, as shown
in Fig.~\ref{fig:approx} for diamond with the same randomly-shifted $k$-point mesh as above.
In Fig.~\ref{fig:approx}(a), we show that the basis set incompleteness error is negligible by 
comparing the spectrum
obtained with two types of pseudopotentials, GTH~\cite{goedecker1996,hartwigsen1998} and ccECP~\cite{bennett2017,bennett2018,annaberdiyev2018,wang2019a}, combined with their
corresponding double- and triple-zeta basis sets. These calculations were performed with a
\mesh{2} $k$-point mesh and without freezing any orbitals.
Additionally, we see that the use of two distinct pseudopotentials 
does not introduce a noticeable difference in the calculated spectra.

In Fig.~\ref{fig:approx}(b), we test the impact of orbital freezing and partitioning
by showing four spectra, all performed with a \mesh{3} $k$-point mesh: 
(1) the EOM-CCSD spectrum without approximations,
(2) the EOM-CCSD spectrum with only
4 occupied and 4 virtual orbitals correlated, 
(3) the partitioned EOM-CCSD spectrum without any frozen orbitals, and 
(4) the partitioned EOM-CCSD spectrum with 4 occupied and 4 virtual orbitals correlated.
We see that freezing orbitals causes a roughly rigid shift of the spectrum to
higher energy by about 0.2--0.5~eV.  The shift is not perfectly rigid and, as
expected, the discrepancy is worst at high energies.  In contrast, the
partitioning error causes a roughly rigid shift to lower energy by a similar
amount. When both approximations are applied, we obtain a spectrum close 
to the one without approximations due to fortuitous cancellation of error, justifying
our use of this affordable approach when scaling up to larger $k$-point meshes. 

The effect of all errors discussed in this subsection can be approximated with a rigid
spectral shift according to the error in the first excitation energy.
These corrections for both $\Gamma$-centered and randomly shifted $k$-point
meshes are summarized in Table~\ref{tab:correction} for all six material studied.
The base result ($E_{555}$) is obtained with partitioned EOM-CCSD using frozen
orbitals and a \mesh{5} $k$-point mesh. 
To this, we apply two composite-style
corrections: $\Delta_{\mathrm{TDL}}$ is the difference between the excitation
energy in the thermodynamic limit obtained by extrapolation and $E_{555}$, where
all calculations are performed with partitioned EOM-CCSD/DZVP with frozen
orbitals, and
$\Delta_{\mathrm{frz+part}}$ is the difference between EOM-CCSD without approximations
and partitioned EOM-CCSD with frozen orbitals, using a \mesh{3} $k$-point mesh.
These two corrections are roughly comparable in magnitude but strongly system dependent.
Although each correction alone may shift the energy by up to 0.5~eV, 
the final correction is typically quite small due to error cancellation.

To reiterate, the final spectra presented in Fig.~\ref{fig:spectra} were obtained with
a \mesh{5} $k$-point mesh using 
partitioned EOM-CCSD, correlating 4 occupied and 4 virtual orbitals per $k$-point,
and then rigidly shifted according to the corrections given in Tab.~\ref{tab:correction}
to approximately correct for finite-size errors, frozen orbitals, and the partitioning
approximation applied to the dense doubles block of the Hamiltonian.

\begin{table}[t]
	\caption{EOM-CCSD first excitation energies and corrections (in eV) for Si, SiC, C, MgO, BN, and LiF.}\label{tab:correction} 
	\begin{ruledtabular}
		\begin{tabular}{l d{-1}  d{-1}  d{-1} d{-1} }		
			\toprule
				& \multicolumn{1}{c}{$E_{555}$}     & \multicolumn{1}{c}{$\Delta_{\mathrm{TDL}}$} & \multicolumn{1}{c}{$\Delta_{\mathrm{frz+part}}$}  & \multicolumn{1}{c}{$E_{\mathrm{final}}$}   \\
				& \multicolumn{4}{c}{randomly shifted $k$-point mesh} \\
				\cline{2-5}
				Si  & 3.52       & -0.18     & 0.20          & 3.53     \\
				SiC & 5.83       & 0.14      & 0.53          & 6.50     \\
				C   & 7.59       & -0.43     & 0.37          & 7.53     \\
				MgO & 8.88       & -0.11     & 0.28          & 9.05     \\
				BN  & 11.06      & -0.16     & 0.24          & 11.14    \\
				LiF & 13.57      & 0.09      & -0.06         & 13.61    \\
				& \multicolumn{4}{c}{$\Gamma$-included $k$-point mesh}   \\
				\cline{2-5}
				Si  & 3.45       & -0.08     & -0.12         & 3.25     \\
				SiC & 6.10       & -0.28     & 0.24          & 6.06     \\
				C   & 7.01       & -0.05     & 0.29          & 7.25     \\
				MgO & 8.03       & 0.20      & 0.41          & 8.64     \\
				BN  & 10.82      & 0.07      & 0.16          & 11.06    \\
				LiF & 13.51      & 0.28      & 0.00          & 13.79   \\
			\bottomrule
		\end{tabular}
	\end{ruledtabular}
\end{table}

\section{Conclusions and outlook}
\label{sec:conc}

We have presented the first absorption spectra of atomistic, three-dimensional
solids using periodic EOM-CCSD, focusing on Si, SiC, C, MgO, BN, and LiF.  With
increasing Brillouin zone sampling, we observe no problems associated with the
lacking size-extensivity of EOM-CCSD spectral intensities~\cite{kobayashi1994,koch1994}.
This may be due to the reasonably complete basis set~\cite{Caricato2009}
provided by a solid-state environment, but further study is warranted. 
After accounting for a number of sources
of error, our best and
final spectra show reasonably good agreement with experimental spectra,
indicating that EOM-CCSD is a promising and tractable approach for the
study of excitations in solids.
In many materials, we find that spectral shapes are well reproduced but are shifted
to higher energies with respect to experiment by about 1~eV. We attribute this
discrepancy to a combination of incomplete electron correlation (i.e., the
impact of triple and higher excitations) and the neglect of zero-point and
finite-temperature vibrational effects~\cite{noffsinger2012,patrick2014,lambrecht2017}.
Unlike in TDDFT and GW-BSE, in CCSD there is reduced freedom in the choice of starting point 
due to its weak sensitivity to the employed reference determinant.

Overall, the agreement between EOM-CCSD and experimental spectra is best for
large-gap insulators and worst for small-gap semiconductors, which we attribute
to finite-size errors, i.e.~incomplete Brillouin zone sampling, and the increasing
importance of correlation in small-gap materials. Whereas
extrapolation of isolated energies to the thermodynamic limit is largely
successful, doing the same for spectral intensities is not straightforward.
The high cost of EOM-CCSD calculations precludes brute force convergence
and future work will explore the use of interpolation~\cite{rohlfing2000}, twist
averaging~\cite{lin2001}, and double-grid schemes~\cite{Kammerlander2012}, which
have been very successful at providing converged GW-BSE spectra at reduced
computational cost.

\section*{acknowledgments}
X.W.~thanks Alan Lewis for helpful discussion.  This work was supported in part
by the National Science Foundation under Grant No.~OAC-1931321.  All
calculations were performed using resources provided by the Flatiron Institute.
The Flatiron Institute is a division of the Simons Foundation.

\end{document}